\documentclass[runningheads]{svmult}

\usepackage{makeidx}   
\usepackage{graphicx}  
\usepackage{subeqnar}  
\usepackage{multicol}  
\usepackage{physprbb}  
\makeindex             

\begin{document}
\title*{The Formation of Stellar Clusters in Turbulent Molecular Clouds:
Effects of the Equation of State}  
%
%
%
%
\titlerunning{Formation of Stellar Clusters}
%
\author{Yuexing Li\inst{1, 2}
\and Ralf S. Klessen\inst{3}
\and Mordecai-Mark Mac Low\inst{1,2}}
\authorrunning{Li, Klessen \& Mac Low}
%
%
\institute{Department of Astronomy, Columbia University, New York, NY
10027, USA  
\and Department of Astrophysics, American Museum of Natural History, New
York
\and Astrophysikalisches Institut Potsdam, D-14482 Potsdam, Germany}

\maketitle              

\begin{abstract}
We study numerically the effect of varying the equation of state (EOS)
on the
formation of stellar clusters in turbulent molecular clouds. Our
results show that the EOS plays an important role in the fragmentation
of the
clouds, and the determination of the initial mass function (IMF) of the
protostellar cores.  Fragmentation decreases with increasing adiabatic
index $\gamma$ in the range $0.2 < \gamma < 1.4$, although the total
amount of mass accreted appears to remain roughly constant through
that range, resulting in more massive cores at higher $\gamma$.
Fragmentation and collapse ceases entirely for $\gamma > 1.4$ as
expected from analytic arguments.  Primordial gas may have effective
$\gamma > 1$, in which case these results may help explain why models of
the formation of the first stars produce isolated, massive objects.
\end{abstract}

\section{INTRODUCTION}
Fragmentation of molecular clouds is key to the formation of star
clusters, yet the conditions for fragmentation are poorly
understood. There have been a few analytical approaches (Jeans 1902;
Penston 1966; Low \& Lynden-Bell 1976), and numerical investigations
of the effects of various physical processes on the collapse, such as
geometry and rotation of the clouds (see Bonnell \& Bastien 1993 for a
review), and magnetic fields (Galli et al. 2001).
Recently, the effects of turbulence have been studied extensively in a
series of 3D simulations (Klessen, Burkert, \& Bate 1998, Paper I;
Klessen, Heitsch, \& Mac Low 2000, Paper II; Heitsch, Mac Low, \&
Klessen 2001, Paper III; and Klessen 2001, Paper IV). However, these
results are based on isothermal models with $\gamma=1$. The effect of
the equation of state (EOS), which is essential in understanding the
physical structure and stability of the turbulent clouds, remains
uncertain.

The balance of heating and cooling in a molecular cloud can be
approximately described by the polytropic EOS, $P = K\rho^{\gamma}$,
where $K$ is a constant, and $P, \rho$ and $\gamma$ are thermal
pressure, gas density and polytropic index, respectively. The density
structures generated by supersonic turbulence depends on the EOS
(Scalo et al. 1998; Passot \& Vazquez-Semadeni 1998; Passot \&
Vazquez-Semadeni 1999). A detailed analysis by Spaans \& Silk (2000)
suggests that $0.2 < \gamma < 1.4$ in the interstellar medium.

In this paper, we perform a 3D numerical survey over this range of
$\gamma$, to investigate the effects of the EOS on the fragmentation
of turbulent clouds and the subsequent formation of a protostellar
cluster.

\section{SIMULATIONS}
\label{sec_result}
We carry out smoothed particle hydrodynamics (SPH) simulations to
determine the effects of EOS on fragmentation.  The code is
implemented with periodic boundary conditions (Klessen 1997) and
``sink'' particles (Bate, Bonnell, \& Price 1995) to replace
high-density cores. The driven turbulence is set up following the $k =
$1--2 model in Mac Low (1999). For detailed descriptions of the code,
see Paper II. The treatment of the polytropic cloud in the code is the
same as in Bonnell \& Bastien (1991). We vary $\gamma$ in steps of 0.1
in otherwise identical simulations.  Self-gravity is included in the
calculation at $t=2.0$ after turbulence is fully established (see
Paper II). Our models use 200,000 particles. The 13 simulations were
performed simultaneously using a serial version of the code on
processors of the Parallel Computing Facility of the AMNH.

\subsection{Fragmentation}
\label{subsec_frag}
Figure 1 shows that the fragmentation properties and the accretion of
the sinks strongly depend on the value of $\gamma$. The lower the
$\gamma$, the earlier fragmentation occurs, and the more fragments
form. It takes longer for gas with high $\gamma$ to fragment. The
slope of the accretion curves are roughly the same, but the number of
sinks is very different for different $\gamma$ at the same time, which
suggests that \textit{individual} protostellar cores is different
for different $\gamma$. In low-$\gamma$ gas, there are more
cores, but the mass of each core is much smaller than
in high-$\gamma$ gas. 

\begin{figure}[tb]
\leavevmode
\includegraphics[height=1.6in]{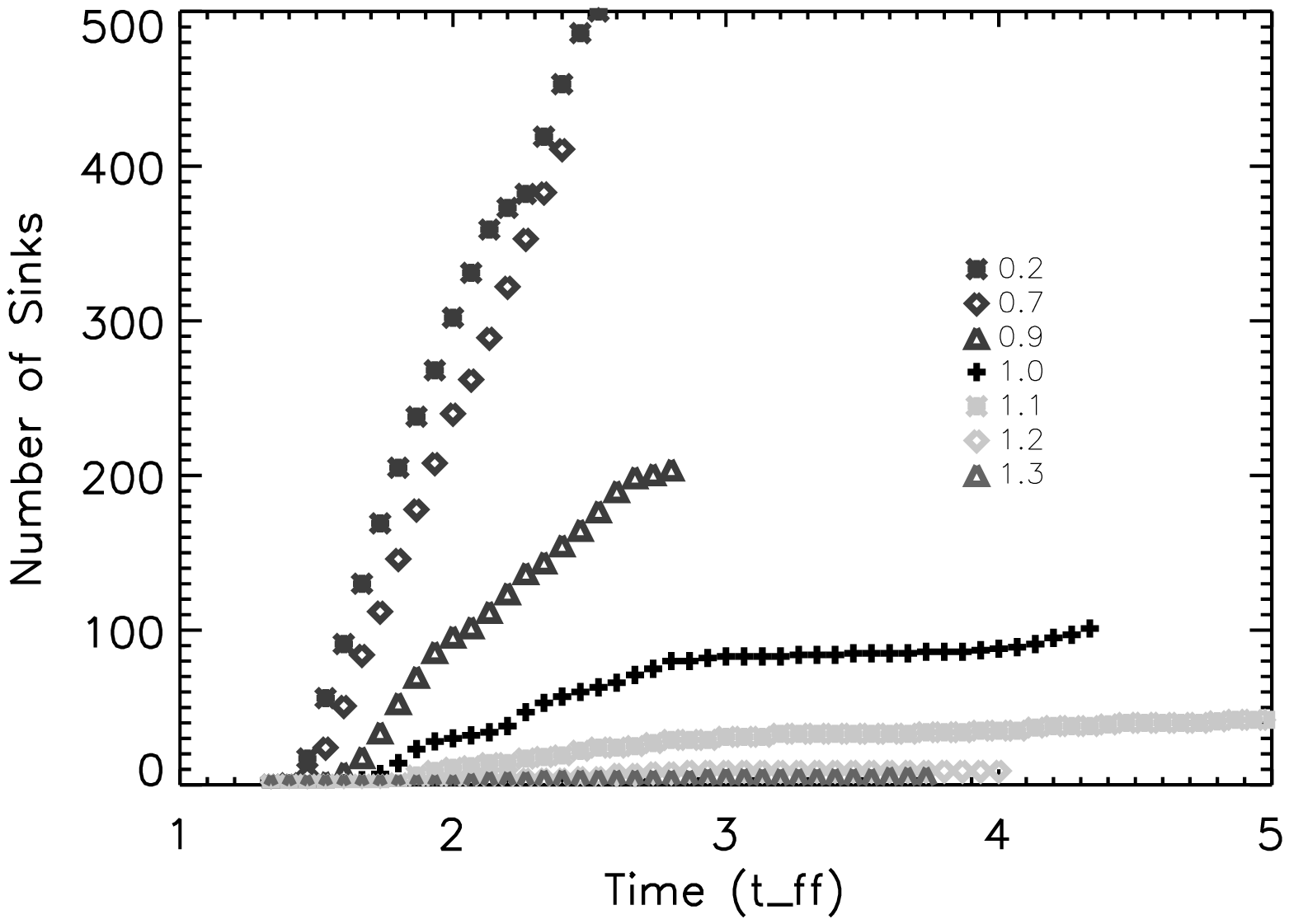}
\includegraphics[height=1.6in]{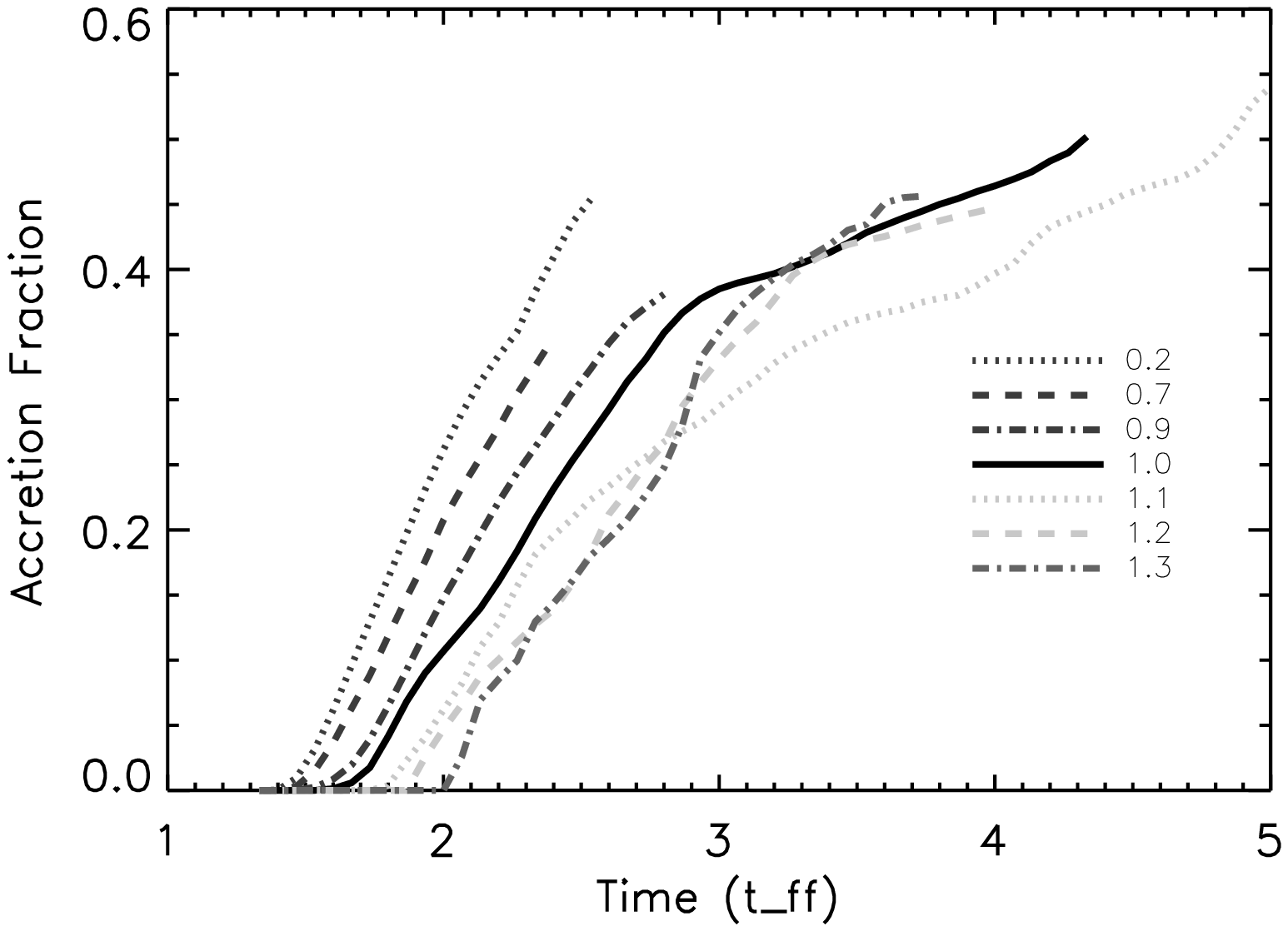}
\caption{\label{fig_sinknum}(\textit{left}) Number of sink particles
representing collapsed star-forming cores as a function of time, and
(\textit{right}) fraction of total mass in the box accreted onto cores
as a function of time, for different $\gamma$.}
\end{figure}

The Jeans mass for a
polytropic gas sphere 
$M_J = (K\pi/G)^\frac32 \gamma^\frac32 \rho^{\frac32(\gamma-\frac43)}$
so 
\begin{equation}
\frac{\partial M_J}{\partial \rho}
\propto \frac{3\gamma - 4}{2}\rho^{(3\gamma - 6)/2},
\end{equation}
which implies that $\gamma = 4/3$ is a critical value during collapse.
When $\gamma < 4/3$, the Jeans mass decreases as density increases,
promoting fragmentation, but when $\gamma > 4/3$, the Jeans mass
\textit{increases} with density, stabilizing collapsing gas.  Indeed,
our model with $\gamma = 1.4$ shows no collapse.

\subsection{The Mass Spectra}
\label{subsec_mas}
Observations show that the mass distribution of gas clumps in various
clouds follows a simple power law, $dN/dM = M^{\nu}$, with $\nu
\approx -1.5$ (Blitz 1993), while the stellar IMF follows $\nu =
-2.33$ at the high-mass end (Salpeter 1955).

\begin{figure}[tb]
\leavevmode
\includegraphics[height=1.2in]{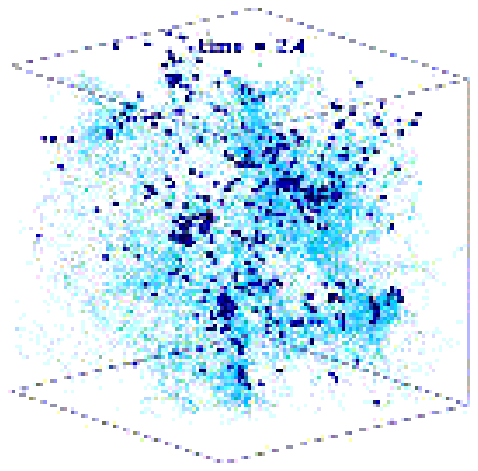}
\includegraphics[height=1.2in]{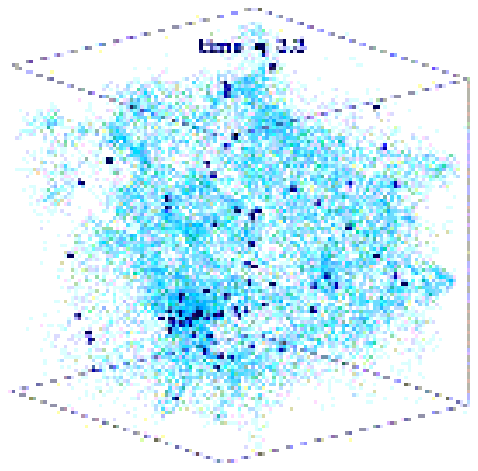}
\includegraphics[height=1.2in]{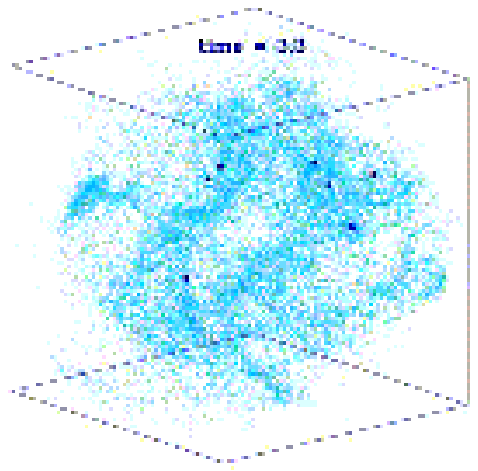}
\includegraphics[height=1.1in]{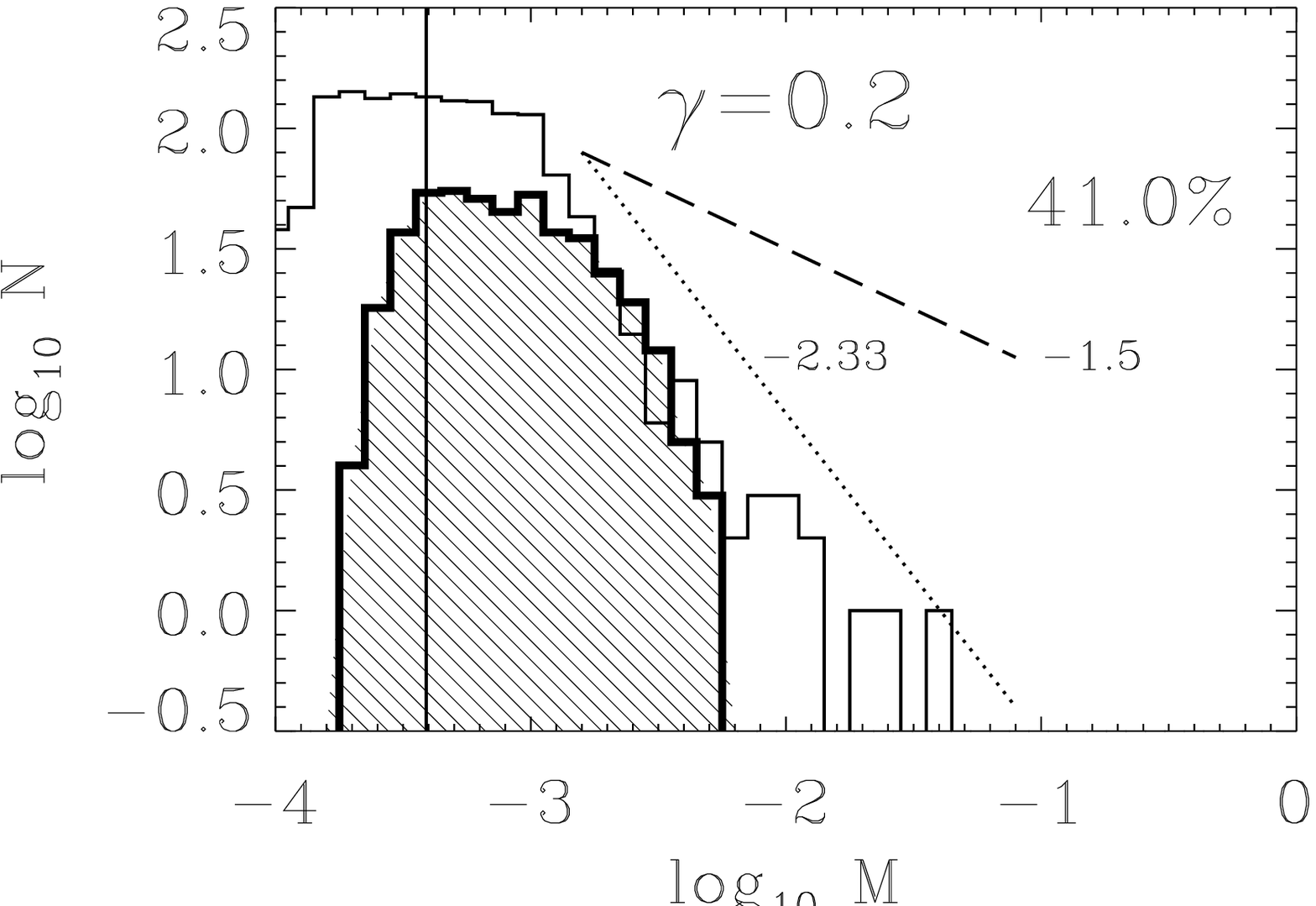}
\includegraphics[height=1.1in]{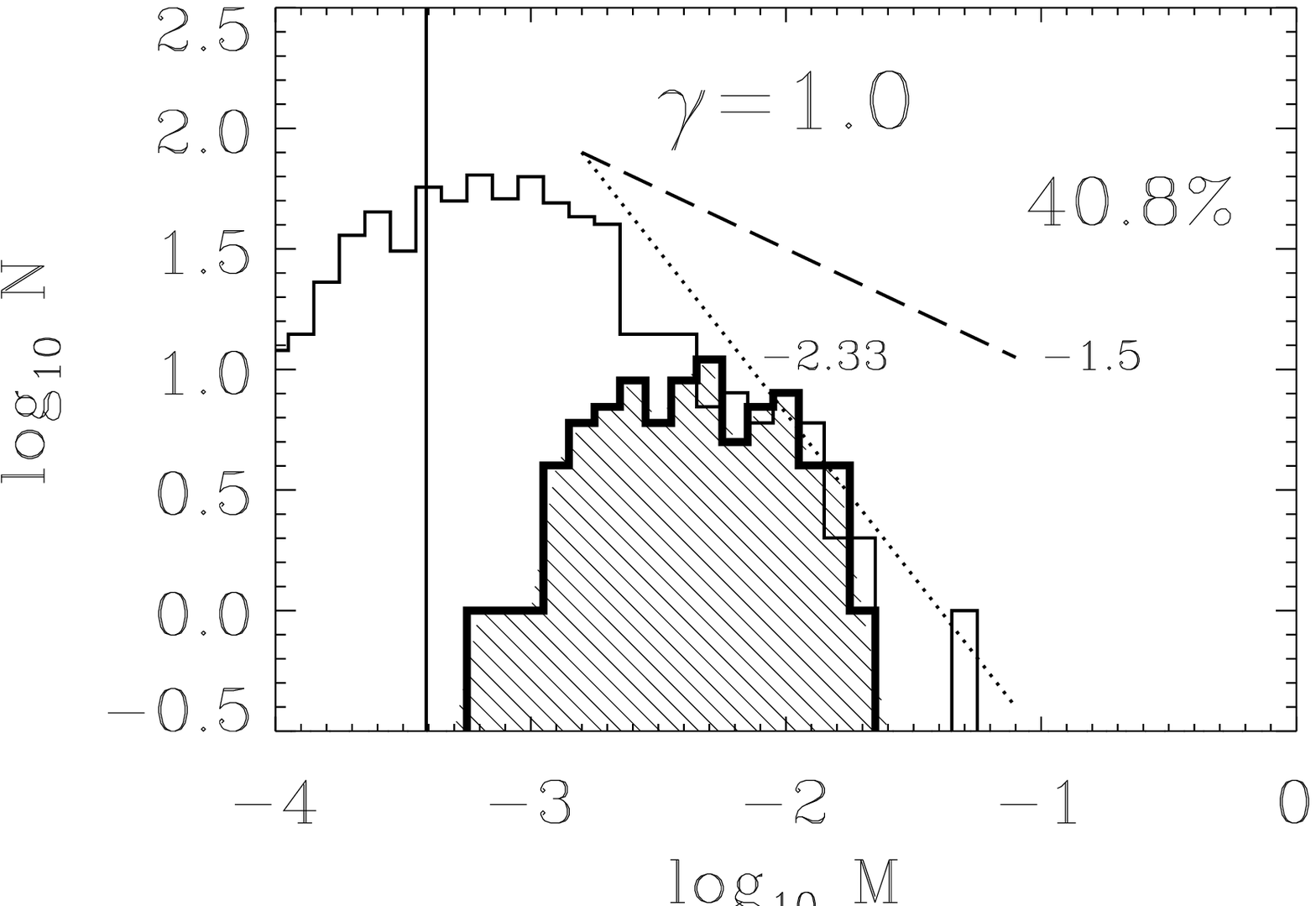}
\includegraphics[height=1.1in]{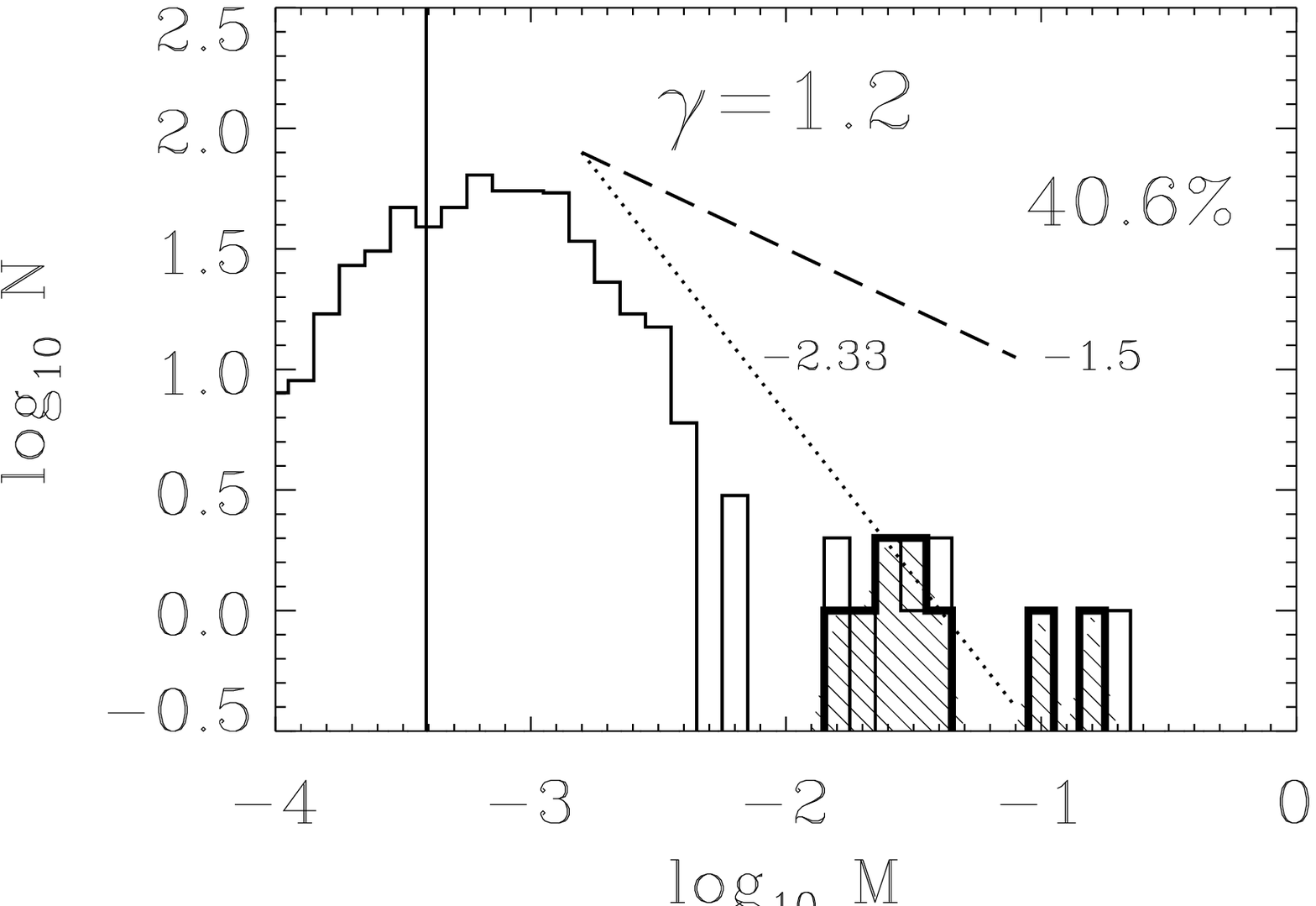}

\caption{\label{fig_mas}Top: 3-D distribution of the gas and protostars for
    different $\gamma$. Bottom: Mass spectra of gas clumps
    (\textit{thin  
    lines}) and of collapsed cores (\textit{hatched thick-lined
    histograms}) for the corresponding cube above. The
    percentage shows the fraction of total mass accreted onto cores. The
    vertical line shows the SPH resolution limit. Shown also are two
    power-law spectra with $\nu = -1.5$ (dashed-line) and $\nu =
    -2.33$ (dotted line).}
\end{figure}
Figure \ref{fig_mas} shows that the spectra of both the clumps and the
cores change with $\gamma$, but that the ones of cores change more
dramatically.  In low-$\gamma$ models, the mass distribution of the
cores at the high-mass end is roughly log-normal.  As $\gamma$
increases, fewer but more massive cores emerge. When $\gamma > 1.0$,
the distribution is dorminated by high mass cores only, and the
spectrum tends to flatten out. It is no longer fitted by either a
log-normal or a power-law. The clump mass spectra, on the other hand,
do show power-law behavior at the high mass side, even for $\gamma >
1.0$.

Our results suggest 
that stars tend to form in clusters in a low-$\gamma$
environment. Protostellar cores are low mass. The apparent lack of
power-law
behavior for the cores in the protostellar cluster might imply that
simple accretion is unable to generate as many high-mass stars as
predicted by the observations, which further suggests that other
mechanisms such as collisions (Bonnell, Bate \& Zinnecker 1998) may be
at work to produce the massive stars in a cluster. Higher resolution
models will be necessary to confirm this, however.

On the other hand, our results also imply that massive stars can form
in small groups or alone in gas with $\gamma > 1.0$.
Spaans \& Silk (2000) suggest that a stiffer EOS ($\gamma > 1.0$) leads
to a peaked IMF, biased toward massive stars, while an EOS with
$\gamma < 1.0$ results in a power-law IMF, in general agreement with our
simulations.

\section{ISOLATED STARS}
\label{sec_discussion}
The formation of isolated massive stars is of great interest, since in
most cases, massive stars are found in clusters. But recently, Lamers
et al. (2002) reported observations of isolated massive
stars or very small groups of a few isolated massive stars in the bulge
of M51. Massey (2002) also reported many field massive stars
in both the Large and Small Magellanic Clouds. From our
simulations, we see that when $\gamma > 1$, only very few or possibly
only one fragment occurs. This cloud core is massive, and would result in the
formation of a high-mass star.

High resolution simulations by Abel, Bryan \& Norman (2002) of the
formation of the first star suggest that initially only one massive
metal-free star forms per pregalactic halo. In the early Universe,
inefficient cooling due to the lack of metals may result in high
$\gamma$.  Our models would then suggest weak fragmentation, resulting
in the formation of only one massive star per cloud.

\vspace{0.1in}{\small We thank M. Fall, H. Lamers, and H. Zinnecker for
valuable discussions, and D. Janies and W. Wheeler for their work on the
Parallel Computing Facility.  YL thanks the AIP for its warm hospitality,
and the Kade Foundation for support of her visit there.  This research
was partially funded by NASA grant NAG5-10103, and by NSF CAREER grant
AST99-85392. }

%

\end{document}